\documentstyle[preprint,tighten,prb,aps]{revtex} 
\begin{document} \draft \preprint{UBCTP-94-001}
\title{Susceptibility of the Spin  1/2 Heisenberg
Antiferromagnetic Chain} \author{Sebastian Eggert$^a$, Ian
Affleck$^{a,b}$ and Minoru Takahashi$^c$}
\address{$^{(a)}$Physics Department and $^b$Canadian Institute
for Advanced Research,}\address{
 University of British Columbia,  Vancouver, B.C.,V6T1Z1,
Canada}\address{  $^c$ Institute for Solid State Physics,
University of Tokyo, Roppongi, Minato-ku, Tokyo 106, Japan}
  \date{\today} \maketitle \begin{abstract}  Highly accurate
results are presented for the susceptibility, $\chi (T)$ of the
$s=1/2$ Heisenberg antiferromagnetic chain for all temperatures,
using the Bethe ansatz and field theory methods.  After going
through a rounded peak, $\chi (T)$ approaches its asympotic
zero-temperature value with infinite slope. 
 \end{abstract} \pacs{75.10.Jm} \narrowtext  In a pioneering
work in 1964, Bonner and Fisher estimated
numerically\cite{Bonner} the susceptibility, $\chi (T)$ for the
$s=1/2$ Heisenberg antiferromagnetic chain, using chain lengths
of up to 11.  The exact value at $T=0$ was  obtained using the
Bethe ansatz\cite{Griffiths,Yang}.  Since that time, the
Bonner-Fisher curve has frequently been used by
experimentalists to determine the value of the exchange
coupling, $J$, and to determine whether or not a given material
has sufficiently small anisotropic exchange and inter-chain
couplings to be approximated by this model.  Recently it has
become possible to calculate this curve much more accurately,
using the Bethe ansatz.\cite{KT}  This method easily gives very
accurate results for $T$ not too small, but becomes
increasingly difficult as $T\to 0$.  On the other hand, an
analytic formula can be derived for $\chi (T)$ at small $T$,
from conformal field theory methods.  Perhaps surprisingly, this
formula predicts, due to the marginally irrelevant operator,
that the susceptibility has infinite slope at $T=0$. Here we
present results for $\chi (T)$  obtained using the Bethe ansatz
for $T\geq .003J$  and compare them to the conformal field
theory prediction, obtaining excellent agreement.  The field
theory prediction for the general $xxz$ model is also given. 
Figure (1) shows the susceptibility obtained from the Bethe
ansatz and Figure (2) compares this result to the field theory
prediction at low $T$.  Note that, with decreasing $T$, after
passing through the maximum, $\chi \approx .147/J$, at
$T\approx .640824J$, the slope of $\chi$ starts to {\it increase}
below the inflection point at $T\approx .087J$, approaching
$\infty$ as $T\to 0$.

The Heisenberg Hamiltonian is written: \begin{equation} H =
J\sum_i \vec S_i\cdot \vec S_{i+1}.\end{equation}  The
susceptibility per site is given by: \begin{equation} \chi (T)
\equiv {1\over T}\sum_i< S^z_iS^z_0>,\end{equation} The low
energy effective field theory description\cite{Affleck1} is
given by the $k=1$ Wess-Zumino-Witten (WZW) non-linear
$\sigma$-model, or equivalently a free, massless boson, with an
effective ``velocity of light'' or spin-wave velocity:
\begin{equation} v=\pi J/2.\end{equation}  This value of $v$ is
determined from the slope of the dispersion relation obtained
from the Bethe ansatz.  The uniform part of the spin density is
given by the conserved current operators, $\vec J_L$, $\vec
J_R$ for left and right-movers: \begin{equation} \vec S_i
\approx \vec J_L(x_i) + \vec J_R(x_i).\end{equation} In the WZW
model $\vec J_L$ and $\vec J_R$ are uncorrelated and their
self-correlations are given by: \begin{eqnarray} <J^a_L(\tau
,x)J^b_L(0,0)>&=&{\delta^{ab}\over 8\pi^2(v\tau
-ix)^2}\nonumber \\
 <J^a_R(\tau ,x)J^b_R(0,0)>&=&{\delta^{ab}\over 8\pi^2(v\tau
+ix)^2}\end{eqnarray} This result can be extended to finite
temperature by a conformal transformation.  This simply
replaces: \begin{equation} v\tau \mp ix \to {v\beta\over
\pi}\sin \left[{(v\tau \mp ix)\pi \over
v\beta}\right],\end{equation} where $\beta \equiv 1/T$.

The susceptibility in the WZW model is thus given by:
\begin{equation} \chi = \int_{-\infty}^\infty dx\left\{
\left[\frac{v\beta}{\pi} \sin \left(\frac{(v\tau
+ix)\pi}{v\beta}\right)\right]^{-2} \ + \ 
\left[\frac{v\beta}{\pi} \sin\left(\frac{(v\tau
-ix)\pi}{v\beta}\right)\right]^{-2} \right\}\end{equation} The
integral can be done by the change of variables: $u =
\tan\frac{\tau \pi}{\beta}$ and  $w=-i\tan\frac{ix
\pi}{v\beta}$, giving: \begin{equation} \chi = {1\over
8\pi^2v}2\pi (1+u^2)\int_{-1}^1{dw\over (u+iw)^2} \ = {1\over
2\pi v}. \label{int}\end{equation} Note that the integral is
independent of $\tau$ or $u$ as it should be since the total
spin is conserved.  Plugging in the spin-wave velocity, $v=J\pi
/2$, gives the zero-temperature susceptibility, $\chi (0) =
1/J\pi^2$.  

The fact that $\chi (T)$ is independent of $T$ in the WZW model
follows from scale invariance.  To obtain the leading
$T$-dependence, we must perturb about the scale invariant fixed
point Hamiltonian with the leading irrelevant operator. This
perturbation is written: \begin{equation} \delta {\cal H} =
{-8\pi^2v\over \sqrt{3}}g\vec J_L\cdot \vec J_R.\end{equation} 
This term is marginally irrelevant for the coupling constant,
$g>0$.  We wish to calculate the correction to $\chi (T)$ to
first order in $\delta {\cal H}$.  Expanding $e^{-\int dxd\tau
\delta {\cal H}(\tau ,x)}$ to first order the correction to
$\chi$ involves four current operators.  Due to the fact that
the left and right currents are uncorrelated, this expression
factorizes into a product of two two-point Green's functions,
one for left-movers and one for right-movers.  Using
translational invariance, the spatial integrals factorize into
two independent integrals of the form of Eq. (\ref{int}),
giving: \begin{equation} \chi = {1\over 2\pi v}+{g\over
v\sqrt{3}}.\end{equation} Again, the correction is naively
temperature-independent, since $g$ is dimensionless.  However,
this formula can be improved by replacing $g$ by $g(T)$, the
effective renormalized coupling at temperature $T$.  By
integrating the lowest order $\beta$ -function, $g(T)$ is given
by:\cite{Affleck2} \begin{equation} g(T) \approx {g_1\over
1+4\pi g_1\ln (T_1/T)/\sqrt{3}}.\end{equation}  Here $g_1$ is
the value of the effective coupling at some temperature $T_1$. 
Both $g_1$ and $g(T)$ must be small for this formula to be
valid.  We may write this more compactly, as: \begin{equation}
g(T) \approx {\sqrt{3}\over 4\pi \ln (T_0/T)}\label{g(T)},\end{equation}
for some temperature $T_0$.  Note that the asympotic behaviour
at small $T$ becomes independent of $g_1$ or $T_0$ up to a
correction of $O((\ln T)^{-2})$. We expect higher order
corrections to $\chi (T)$ of $O(g(T)^2)$, $O(g(T)^3)$, etc.  By
shifting $T_0$, we can eliminate the $O(\ln T)^{-2})$ part of
the $O(g(T)^2)$ correction. Thus we obtain the leading
$T$-dependence of $\chi$: \begin{equation} \chi (T) = {1\over
2\pi v}+ {1\over 4\pi v\ln (T_0/T)} + O((\ln
T)^{-3}).\label{chiv}\end{equation} This formula is universal
in the sense that if we add additional interactions to the
Heisenberg Hamiltonian that respect the $SU(2)$ and
translational symmetry and are not sufficiently large as to
drive it into another phase, then this formula applies for some
values of $v$ and $T_0$. In particular, if we add an
antiferromagnetic second nearest neighbour interaction, $J_2$,
the value of the bare coupling, $g$, decreases and reaches zero
at $J_2\approx .24J$.  At this point all logarithmic terms in
$\chi (T)$ vanish, corresponding to $T_0 \to \infty$ and $\chi
(T)$ should be analytic near $T=0$.  For larger $J_2$ the
system develops a gap and $\chi (T)$ vanishes exponentially as
$T\to 0$. Eq. (\ref{chiv}) should be valid for arbitrary
half-integer spin Heisenberg antiferromagnets. For the ordinary
$s=1/2$ Heisenberg model, $v=J\pi /2$ giving: \begin{equation}
J\pi^2\chi (T) \approx 1 + {1\over 2\ln
(T_0/T)}.\label{chi}\end{equation} A good fit to the Bethe
ansatz data is obtained with $T_0\approx 7.7 J$, as  shown in
Figure (2).  Eq. (\ref{chi}) is valid to within 2\% for $T<.1
J$. Similar formulas for the finite-size dependent of low
energy states were obtained in Ref. (\onlinecite{Affleck2})
with $T_0/T$ replaced by $L/L_0$ where $L$ is the size of the
system.  The role of
infrared cut-off on the renormalization of the coupling
constant is played by either the length, $L$ or an effective
thermal length $v/T$ in the two cases.  Note that the field
theory calculation of the  susceptibility is done in the
limit $LT/v \to \infty$ whereas the finite-size spectrum
calculations are done in the opposite limit $LT/v \to 0$.  In
both cases the space-imaginary time surface is an infinite
length cylinder of circumference $v/T$ or $L$ respectively.
By pushing all calculations to
second order in $g$, predictions could be made relating the
various values of the $L_0$'s for different energy levels and
$T_0$. Alternatively, we may use the deviation of $\chi (T)$
and the energy levels from their asympototic values to define
four different estimates of the effective coupling constant
versus length.  These are shown in Figure 3.  The singlet and
triplet excited state estimates of the effective coupling,
$g(L)$, only converge very slowly as $g(L)^2$, as we expect
since, in general, all these quantities receive corrections at
next order in $g$.  Surprisingly, the susceptibility and
triplet estimates of $g$, using the relationship $L
\leftrightarrow v/T$, appear to converge much more rapidly as
$1/L^2$, suggesting the absence of corrections to any finite
order in $g(L)$.  [The groundstate and triplet estimates appear
to converge rapidly to a small non-zero difference.  This
remains a puzzlingly descrepancy between conformal field theory
and Bethe ansatz calculations.  See Ref. (\onlinecite{Nomura})
for further discussion.]

We now consider the general $xxz$ model: \begin{equation} H =
J\sum_i\left[ S_i^xS_{i+1}^x+S_i^yS_{i+1}^y+\Delta
S_i^zS_{i+1}^z\right].\end{equation}  For $\Delta >1$, the
Hamiltonian has a N\'eel ordered groundstate and a gap.  Hence
the susceptibility vanishes exponentially at low $T$.  For
$\Delta <1$, the system remains gapless.  It is now convenient
to use abelian bosonization, involving a free boson $\phi$. 
The $zz$ component becomes: \begin{equation} J^z_LJ^z_R =
{1\over 4\pi} \left[\left({1\over v}{\partial \phi \over
\partial t}\right)^2-\left({\partial \phi \over \partial
x}\right)^2\right].\end{equation}  The other components of the
interaction become: \begin{equation} J^x_LJ^x_R+J^y_LJ^y_R
\propto \cos \sqrt{8\pi}\phi .\end{equation}  The $zz$ part is
exactly marginal and has the effect of rescaling the
boson:\cite{Affleck1} \begin{equation} \phi \to {\phi \over
\sqrt{2\pi}R},\end{equation} with $R<1/\sqrt{2\pi}$.  The other
part then becomes: \begin{equation} J^x_LJ^x_R+J^y_LJ^y_R
\propto \cos (2\phi /R) .\end{equation}  This has scaling
dimension  \begin{equation} x=1/\pi R^2>2 \end{equation} and is
irrelevant.  It is the leading irrelevant operator, provided
$x<4$.
 A renormalization of the velocity also occurs. The zero
temperature susceptibility gets rescaled to: \begin{equation}
\chi (0) = {1\over v(2\pi R)^2}.\end{equation}  The first order
contribution to the susceptibility from $\cos (2\phi /R)$
vanishes.  The second order contribution is determined by a
standard scaling argument, giving:
 \begin{equation} \chi (T) \to {1\over v(2\pi R)^2}\ +\
\hbox{constant}\cdot T^{(2/\pi R^2-4)}
.\label{chianis}\end{equation} Again this formula is universal
in the sense that the two parameters $v$ and $R$ determine all
low energy features of the model and $R<1/\sqrt{2\pi}$. 
Increasing anisotropy leads to decreasing $R$. Eq.
(\ref{chianis}) should apply for arbitrary half-integer spin
$xxz$ antiferromagnets in the gapless phase, provided that
$1/\pi R^2<3$. Otherwise the exponent is replaced by $2$. For
the $s=1/2$ Heisenberg model $R(\Delta )$ and $v(\Delta )$ have
been determined from the Bethe
ansatz.\cite{Baxter,Affleck1,desCloiseaux} Letting:
\begin{equation} \Delta = \cos \theta \end{equation} we find:
\begin{eqnarray} \sqrt{2\pi}R(\Delta )&=&\sqrt{1-{\theta \over
\pi}}\nonumber \\ v(\Delta ) &=& {J\pi \sin \theta \over
2\theta}.\end{eqnarray}
 Thus  \begin{equation} \chi (0) ={\theta \over \pi (\pi
-\theta )\sin \theta}\end{equation} (in agreement with the
explicit Bethe ansatz calculation\cite{Yang}) and the exponent
is given by: \begin{equation} 2 /\pi R^2-4={4 \theta \over \pi
- \theta}.\end{equation} For small anisotropy: \begin{equation}
2\pi /R^2-4 \approx {4\sqrt{2(1-\Delta )}\over
\pi}.\end{equation} $\chi (T)$ has an infinite slope at $T=0$
for $\Delta >\cos^{-1}(\pi /5)\approx .809$.  For $\Delta <.5$,
the exponent $(2\pi /R^2-4)$ is replaced by $2$.  

These results may prove useful in experimental studies of
quasi-one-dimensional antiferromagnets.  In particular, we may
define a ``critical region'' where the temperature dependence
is governed by the leading irrelevant operator so that Eqs.
(\ref{chiv}) or (\ref{chianis}) hold.  For the isotropic
Heisenberg model, this critical region can be identified
approximately as $T < .1 J$, below the inflection point.  Only
if the N\'eel temperature, determined by interchain couplings,
is below this value can the one dimensional critical behaviour
be observed.

\section{Acknowledgements} We would like to thank V. Korepin
for a helpful correspondence.  This research is supported in
part by NSERC of Canada and by
Grants-in-Aid for Scientific Research on Priority Areas,
"Computational Physics as a New Frontier in Condensed Matter
Research" (Area No 217) and "Molecular Magnetism" (Area No 228),
from the Ministry of Education, Science and Culture, Japan.

\begin{figure} \caption{$\chi (T)$ from the Bethe
ansatz.  $\chi (0)$ is taken from Ref.
(\protect{\onlinecite{Griffiths}}).} \end{figure} \begin{figure}
\caption{Field theory [Eq. (\protect{\ref{chi}})] versus Bethe
ansatz formulas for $\chi (T)$ at low temperature.}\end{figure}
\begin{figure}\caption{Estimates of the effective coupling from
lowest order perturbation theory correction to finite-size
energy of groundstate, first excited triplet state, first
excited singlet state and to susceptibility, using
$L\leftrightarrow v/T$.  The renormalization group prediction of
Eq. (\protect{\ref{g(T)}}) is also shown.} \end{figure}
 \end{document}